# A Comparison of Accuracy for Methods to Approximate Fisher Information

Shenghan Guo

**Department of Applied Mathematics and Statistics**
**The Johns Hopkins University**
**Baltimore, Maryland  21218-2682**

December 2014

# Abstract

The Fisher information matrix (FIM) has long been of interest in statistics and other areas. It is widely used to measure the amount of information in a set of data and to calculate the lower bound (*Cramér−Rao bound*) of the variance for estimates such as maximum likelihood and to conduct score tests. (It is also of interest to note that other measures of information in the data are useful in practice, including the Kullback Liebler-Lindley measure, which is derived from the information theoretic notions of Shannon information and the distance between two probability distributions.) In practice, we do not always know the actual FIM. This is often because obtaining the first or second-order derivatives of the log-likelihood function is difficult, or simply because the calculation of FIM is too formidable. In such cases, we need to use the approximation of FIM. In general, there are two ways to estimate FIM. One is to use the product of gradient and the transpose of itself, and the other is to calculate the Hessian matrix and then take negative sign. Mostly people use the latter method in practice. However, this is not necessarily the better way. To find out which of the two methods is better, we need to conduct a theoretical study to compare their efficiency. In this paper we mainly focus on the case where the unknown parameter that needs to be estimated is scalar and the random variables we have are independent. In this scenario, Fisher information matrix is virtually Fisher information number (FIN). Using the Central Limit Theorem (CLT), we get asymptotic variances for the two methods, by which we compare their accuracy. Taylor expansion assists in estimating the two asymptotic variances. A numerical study is provided as an illustration of the conclusion. The next is a summary of limitations of this paper. We also enumerate several fields of interest for future study in the end of this paper.

# 1. Introduction

Given the importance and numerous applications of the Fisher information matrix (FIM), it is natural to think of the calculation of this quantity. Nonetheless, as mentioned, calculation of FIM is not always practical. Sometimes estimation or approximation makes more sense when feasibility and cost (in all respects) are taken into account. Basically, situations in which we need to resort to approximation include (1) the analytical forms of the gradient and the Hessian matrix (or the Hessian number when the unknown parameter is scalar) are not accessible, nor is any directly usable values, and (2) analytical computation of FIN is extremely troublesome that an unacceptably high cost is resulted.

There are many examples of estimation problems where it is difficult to get FIM. Let us mention several here. In Spall and Garner (1990), when using a state-space model with nuisance parameter, FIM is needed to provide a lower bound for the variance of the maximum likelihood estimator. However, the influence of nuisance parameters makes it difficult to get the real value of FIM. Spall, J. C. and Maryak, J. L. (1992) provides a feasible Bayesian estimator for non-i.i.d. projectile measurements and quintiles, in which the approximation of FIM is necessary. McLachlan, G. J. and Peel, D. (2000), Sect. 2.15 gives an example where we need to use the approximation of FIM instead of analytical calculation, which is when finite mixture models are involved. Analytically calculating FIM (or Fisher information number for a scalar unknown parameter) for mixture density functions could be costly in all respects. In Favennec, Y. (2007), it is also preferable to use the approximation of FIM since inverse heat condition problems bring challenges to the analytical calculation. A system with binary subsystems is described in Spall (2014), for which the approximation of FIM becomes the optimal option because the complexity of the system makes it almost impossible to provide an analytical calculated result without considerable efforts.

In the aforementioned cases, approximation rather than analytical calculation is usually the optimal strategy. This raises the question that what kind of existing methods give the best approximation to FIM.

According to the definition, when the unknown parameter $\theta$ is scalar, the Fisher information matrix becomes the Fisher information number (FIN). We can calculate FIN in two ways. With a likelihood function $l(\theta | \mathbf{Z}_n) = p(\mathbf{Z}_n | \theta)$, where $p(\mathbf{Z}_n | \theta) = \prod_{i=1}^{n} p_i(z_i | \theta)$ is the density function of a series of independent random variable $\mathbf{Z}_n = [z_1, z_2, ..., z_n]^T$ and depends on the parameter $\theta$, we have the gradient

$$g(\theta | \mathbf{Z}_n) = \frac{\partial \log l(\theta | \mathbf{Z}_n)}{\partial \theta}, \tag{1.1}$$

and

$$H(\theta | \mathbf{Z}_n) = \frac{\partial^2 \log l(\theta | \mathbf{Z}_n)}{\partial \theta^2}. \tag{1.2}$$

Denote FIN by $F_n(\theta)$, then FIN is defined to be

$$F_n(\theta) = E\left[g^2(\theta | \mathbf{Z}_n)\right], \tag{1.3}$$

which, under standard conditions (e.g., Spall, 2003, pp. 352–353), is equal to

$$F_n(\theta) = -E\left[H(\theta | \mathbf{Z}_n)\right]. \tag{1.4}$$

In practice, people use estimation when the true FIN is inaccessible. Let us define

$$\bar{G}_n(\theta) = \frac{1}{n}\sum_{i=1}^{n} g_i^2(\theta | z_i) = \frac{1}{n}\sum_{i=1}^{n} g_i^2(z_i), \tag{1.5}$$

$$\bar{H}_n(\theta) = -\frac{1}{n}\sum_{i=1}^{n} H_i(\theta | z_i) = -\frac{1}{n}\sum_{i=1}^{n} H_i(z_i), \tag{1.6}$$

(1.5) and (1.6) are two unbiased estimators for the averaged FIN, i.e. $F_n(\theta)/n$.

There are no strict settings for when to use which form. In many settings, people tend to use (1.6), though it appears that there is no theoretical support showing that the Hessian matrix gives a better approximation to the actual FIN. Such reality is the direct motivation for this project. We would like to know which method gives the better estimation. If we have the answer to this inquiry, we are able to choose the optimal way to approximate FIN in different situations, which leads to more desirable results.

Studies involved in FIN are common, but works that directly tackle our problem are not easily found. Most related works concentrate on the comparison of the expected FIN and the observed FIN. Efron and Hinkley (1978) look at the accuracy of these two with regard to approximating variance for the MLE estimator. They come up with an assertion that for the scalar case, the inverse of the observed FIN gives a better estimation with a well-chosen auxiliary statistic $a$. Cao and Spall (2009) discuss a similar issue with Efron and Hinkley (1978). For an unknown target parameter $\theta$ and its MLE $\hat{\theta}_n$, Cao and Spall (2009) adjust the estimation target to

$$\min_{T(\mathbf{X})} E\left[\left(n \operatorname{var}(\hat{\theta}_n) - T(\mathbf{X})\right)^2\right], \tag{1.7}$$

with $T(\mathbf{X})$ being an estimate of the variance of normalized $\hat{\theta}_n$ given sample $\mathbf{X}$. For this revised problem, Cao and Spall (2009) show an opposite conclusion that the inverse of the expected FIN is preferable under certain conditions. The dissertation of Xumeng Cao (2013) extends this conclusion to multivariate $\boldsymbol{\theta} = [t_1, t_2, ..., t_n]^T$ (i.e., to the FIM, not just FIN, case).

The above works are to some extent relevant to our interest yet they do not lead to a straightforward conclusion for us. Nonetheless, Spall (2005) comes up with a new Monte Carlo method to estimate FIM and gives some theoretical analysis concerning the efficiency of the Hessian matrix form and the product-of-gradient form in approximating the true FIM. It is shown that under certain circumstances the Hessian matrix form is a better choice. This assertion, as well as the reasoning process, provides some clues for our study. Specifically, we use Taylor expansion to analyze the asymptotic variances just as Spall (2005) does, which enables us to simplify the calculation and leads to a relatively succinct result.

## 2. Background

In this section, we provide some background knowledge that we use in this paper, in order to help the reader get a clear picture of our study.

### 2.1 The Central Limit Theorem

We can apply CLT to (1.5) and (1.6) and assert that they are asymptotically normally distributed. Hence we can analyze their distributions and find out the accuracy of (1.5) and (1.6).

The classic CLT applies to independent and identically distributed (i.i.d.) samples.

**Lindeberg–Lévy CLT**:

Suppose $(X_1, X_2, ...)$ is a sequence of i.i.d. random samples with finite mean and variance $\sigma^2$. Define $S_n = n^{-1} \sum_{i=1}^{n} X_i$, where $n$ is the sample size. As $n$ approaches infinity, the random variable $\sqrt{n}(S_n - \mu)$ converges in distribution to a normal random variable with mean 0 and variance $\sigma^2$. That is,

$$\sqrt{n}(S_n - \mu) \xrightarrow{dist} N(0, \sigma^2) \tag{2.1}$$

There are other versions of CLT that apply to independent and non-identically distributed (i.n.i.d.) samples.

**Lyapunov CLT:**

Suppose $(X_1, X_2, ...)$ is a sequence of i.n.i.d. random samples with finite mean $\mu_i$ and variance $\sigma_i^2$.

Define $s_n^2 = \sum_{i=1}^n \sigma_i^2$, where $n$ is the sample size. If for some $\delta > 0$, the *Lyapunov's condition*

$$\lim_{n \to \infty} \frac{1}{s_n^{2+\delta}} \sum_{i=1}^n E\left[|X_i - \mu_i|^{2+\delta}\right] = 0. \tag{2.2}$$

is satisfied, then as $n$ approaches infinity, the random variable $s_n^{-1} \sum_{i=1}^n (X_i - \mu_i)$ converges in distribution to a standard normal random variable,

$$\frac{1}{s_n} \sum_{i=1}^n (X_i - \mu_i) \xrightarrow{dist} N(0,1). \tag{2.3}$$

**Lindeberg CLT:**

Lindeberg CLT is built in the same setting with Lyapunov CLT, with only the *Lyapunov's condition* replaced by the *Lindeberg's condition*, which is much weaker. The *Lindeberg's condition* says that, for every $\varepsilon > 0$, if

$$\lim_{n \to \infty} \frac{1}{s_n^2} \sum_{i=1}^n E\left[(X_i - \mu_i)^2 \cdot \mathbf{1}_{\{|X_i - \mu_i| > \varepsilon s_n\}}\right] = 0, \tag{2.4}$$

where $\mathbf{1}_{\{\ldots\}}$ is an indicator, then the result from the Lyapunov CLT holds:

$$\frac{1}{s_n} \sum_{i=1}^n (X_i - \mu_i) \xrightarrow{dist} N(0,1).$$

**2.2 Taylor expansion (Taylor series)**

Taylor expansion is very useful for approximating values that are difficult to calculate directly. In this paper, we use the second-order Taylor expansion to estimate the asymptotic variances of (1.5) and (1.6).

According to Rudin W. (1976), the Taylor expansion of a function $f(x)$ that is infinitely differentiable at some point $a$ is

$$f(a) + \frac{f'(a)}{1!}(x-a) + \frac{f''(a)}{2!}(x-a)^2 + \frac{f^{(3)}(a)}{3!}(x-a)^3 + \cdots. \tag{2.5}$$

where $f'(\cdot)$ denotes the first-order derivative of $f$ with respect to the variable $x$, $f''(\cdot)$ denotes the second-order derivative of $f$ with respect to the variable $x$ and $f^{(n)}(\cdot)$ denotes the $n$th derivative of $f$ with respect to the variable $x$.

Theoretically, we could expand a function till $n! f^{(n)}(\cdot)(x-a)^n$ with $n$ sufficiently large. However, when using Taylor expansion, people often only expand the function to its second-order derivative and ignore the remaining items that could possibly appear in its Taylor series. This is because the denominator $n!$ in $f^{(n)}(\cdot)(x-a)^n/n!$ increases very fast and items with higher order become insignificant quickly. Based on this rational, we only use second-order Taylor expansion in the following context. Also, we assume that the density functions we consider are at least twice-differentiable with respect to $\theta$.

## 3. Theoretical Analysis

In this section, we present the theoretical analysis on the accuracy of (1.5) and (1.6).

For $n$ independent random variables $\mathbf{Z}_n = [z_1, z_2, ..., z_n]^T$, the likelihood function is

$$l(\theta | \mathbf{Z}_n) = p(\mathbf{Z}_n | \theta) = \prod_{i=1}^{n} p_i(z_i | \theta), \tag{3.1}$$

where $p_i(\cdot | \theta)$ is the density function for random variable $z_i$ that depends on the parameter $\theta$, $i = 1, 2, ..., n$.

The gradient is the first-order derivative of the log-likelihood loss function with respect to $\theta$,

$$g(\theta | \mathbf{Z}_n) = \frac{\partial \log l(\theta | \mathbf{Z}_n)}{\partial \theta} = \frac{\partial}{\partial \theta}\left(\sum_{i=1}^{n} \log p_i(z_i | \theta)\right) = \sum_{i=1}^{n} \frac{\partial \log p_i(z_i | \theta)}{\partial \theta}. \tag{3.2}$$

The Hessian matrix turns out to be

$$H(\theta | \mathbf{Z}_n) = \frac{\partial^2 \log l(\theta | \mathbf{Z}_n)}{\partial \theta^2} = \frac{\partial}{\partial \theta}\left(\sum_{i=1}^{n} \frac{\partial \log p_i(z_i | \theta)}{\partial \theta}\right) = \sum_{i=1}^{n} \frac{\partial^2 \log p_i(z_i | \theta)}{\partial \theta^2}. \tag{3.3}$$

According to the definition of FIN, we have

$$\begin{aligned} F_n(\theta) &= E\left[g^2(\theta | \mathbf{Z}_n)\right] \\ &= E\left[\left(\sum_{i=1}^{n} \frac{\partial \log p_i(z_i | \theta)}{\partial \theta}\right)^2\right] \\ &= E\left[\sum_{i=1}^{n}\left(\frac{\partial \log p_i(z_i | \theta)}{\partial \theta}\right)^2 + 2\sum_{j<k} \frac{\partial \log p_j(z_j | \theta)}{\partial \theta}\frac{\partial \log p_k(z_k | \theta)}{\partial \theta}\right] \\ &= \sum_{i=1}^{n} E\left[\left(\frac{\partial \log p_i(z_i | \theta)}{\partial \theta}\right)^2\right] + 2\sum_{j<k} \underbrace{E\left[\frac{\partial \log p_j(z_j | \theta)}{\partial \theta}\frac{\partial \log p_k(z_k | \theta)}{\partial \theta}\right]}_{(*)}. \end{aligned} \tag{3.4}$$

Since $z_j$ and $z_k$ are independent for $j \neq k$, $j, k = 1, 2, ..., n$,

$$(*) = E\left[\frac{\partial \log p_j(z_j \mid \theta)}{\partial \theta} \frac{\partial \log p_k(z_k \mid \theta)}{\partial \theta}\right]$$

$$= E\left[\frac{\partial \log p_j(z_j \mid \theta)}{\partial \theta}\right] E\left[\frac{\partial \log p_k(z_k \mid \theta)}{\partial \theta}\right]. \tag{3.5}$$

Assume that the Dominated Convergence Theorem (DCT) applies, so that integration and differentiation can be interchanged. Then

$$E\left[\frac{\partial \log p(z \mid \theta)}{\partial \theta}\right] = \int_{-\infty}^{+\infty} \frac{\partial \log p(z \mid \theta)}{\partial \theta} p(z \mid \theta) dz$$

$$= \int_{-\infty}^{+\infty} \frac{1}{p(z \mid \theta)} \frac{\partial p(z \mid \theta)}{\partial \theta} p(z \mid \theta) dz$$

$$= \int_{-\infty}^{+\infty} \frac{\partial p(z \mid \theta)}{\partial \theta} dz = \frac{\partial}{\partial \theta}\left(\int_{-\infty}^{+\infty} p(z \mid \theta) dz\right)$$

$$= \frac{\partial}{\partial \theta}(\text{constant}) = 0. \tag{3.6}$$

Therefore, we have that expansion $(*) = 0$. So

$$F_n(\theta) = \sum_{i=1}^{n} E\left[\left(\frac{\partial \log p_i(z_i \mid \theta)}{\partial \theta}\right)^2\right] = \sum_{i=1}^{n} E\left[g_i^2(z_i)\right]. \tag{3.7}$$

With the alternative definition of FIN,

$$F_n(\theta) = -E[H(\theta \mid \mathbf{z})] = -E\left[\sum_{i=1}^{n} \frac{\partial^2 \log p_i(z_i \mid \theta)}{\partial \theta^2}\right]$$

$$= -\sum_{i=1}^{n} E\left[\frac{\partial^2 \log p_i(z_i \mid \theta)}{\partial \theta^2}\right] = -\sum_{i=1}^{n} E[H_i(z_i)], \tag{3.8}$$

**Scenario A. Independent and identically distributed (i.i.d.) samples**

In the most ideal case where we have i.i.d. samples, all of which have the same density function $p(\cdot \mid \theta)$. Thus the averaged FIN, $F_n(\theta)/n$, turns out to be either $E[g(z_i)]$ or $-E[H(z_i)]$ for any $z_i$ in $\mathbf{Z}_n = [z_1, z_2, ..., z_n]^T$, with the subscript $i$ in $g_i(\cdot)$ and $H_i(\cdot)$ ignored because of equality in density functions.

According to the CLT, we have the following fact:

$$\sqrt{n}\left(\bar{G}_n(\theta) - F(\theta)/n\right) \xrightarrow{dist} N\left(0, \mathrm{var}\left[g^2(z_i)\right]\right), \tag{3.9}$$

$$\sqrt{n}\left(\bar{H}_n(\theta) - F(\theta)/n\right) \xrightarrow{dist} N\left(0, \mathrm{var}\left[H(z_i)\right]\right). \tag{3.10}$$

Now what we need to do is to contrast the magnitude of $\text{var}\left[g^2(z_i)\right]$ and $\text{var}\left[H(z_i)\right]$. The expression in either (1.5) or (1.6) that has the smaller variance is therefore the better choice for FIN estimation. Since the calculation of $\text{var}\left[g^2(z_i)\right]$ and $\text{var}\left[H(z_i)\right]$ is challenging, we turn to the second-order Taylor expansion to give a close approximation to these two. Define $\mu = E(z_i)$. Let

$$g'(\mu) = \left.\frac{\partial g(z_i)}{\partial z_i}\right|_{z_i=\mu} \quad \text{and} \quad g''(\mu) = \left.\frac{\partial^2 g(z_i)}{\partial z_i^2}\right|_{z_i=\mu} ;$$

$$H'(\mu) = \left.\frac{\partial H(z_i)}{\partial z_i}\right|_{z_i=\mu} \quad \text{and} \quad H''(\mu) = \left.\frac{\partial^2 H(z_i)}{\partial z_i^2}\right|_{z_i=\mu} ,$$

where $i = 1, 2, \ldots, n$. The second-order expansions for the functions $g^2(\cdot)$ and $H(\cdot)$ are shown in the Appendix. Using the expansions, the variances are

$$\begin{aligned}
\text{var}\left[g^2(z_i)\right] &\approx 4g^2(\mu)[g'(\mu)]^2 \sigma^2 + \left([g'(\mu)]^2 + g(\mu)g''(\mu)\right)^2 \text{var}\left[(z_i - \mu)^2\right] \\
&+ \frac{1}{16}[g''(\mu)]^4 \text{var}\left[(z_i - \mu)^4\right] + [g'(\mu)]^2[g''(\mu)]^2 \text{var}\left[(z_i - \mu)^3\right] \\
&+ 4g(\mu)g'(\mu)\left([g'(\mu)]^2 + g(\mu)g''(\mu)\right) E\left[(z_i - \mu)^3\right] \\
&+ g(\mu)g'(\mu)[g''(\mu)]^2 E\left[(z_i - \mu)^5\right] + 4g(\mu)[g'(\mu)]^2 g''(\mu) E\left[(z_i - \mu)^4\right] \\
&+ \frac{1}{2}[g''(\mu)]^2 \left([g'(\mu)]^2 + g(\mu)g''(\mu)\right)\left(E\left[(z_i - \mu)^6\right] - \sigma^2 E\left[(z_i - \mu)^4\right]\right) \\
&+ 2g'(\mu)g''(\mu)\left([g'(\mu)]^2 + g(\mu)g''(\mu)\right)\left(E\left[(z_i - \mu)^5\right] - \sigma^2 E\left[(z_i - \mu)^3\right]\right) \\
&+ \frac{1}{2}g'(\mu)[g''(\mu)]^3 \left(E\left[(z_i - \mu)^7\right] - E\left[(z_i - \mu)^4\right] E\left[(z_i - \mu)^3\right]\right),
\end{aligned} \qquad (3.11)$$

and

$$\text{var}\left[H(z_i)\right] \approx [H'(\mu)]^2 \sigma^2 + \frac{1}{4}[H''(\mu)]^2 \text{var}\left[(z_i - \mu)^2\right] + H'(\mu)H''(\mu) E\left\{[z_i - \mu]^3\right\}. \qquad (3.12)$$

Detailed calculations can be found in the Appendix.

The above results may seem intimidating at the first glance, especially $\text{var}\left[g^2(z_i)\right]$, whereas they are not complicated in nature. Let's consider a most common type of density functions, symmetric density. A basic fact about such density functions is that the odd central moments are 0. Applying our results to this kind of density, we are able to drop off a couple of items, giving much more succinct forms.

$$\text{var}\left[g^2(z_i)\right] \approx 4g^2(\mu)[g'(\mu)]^2\sigma^2 + \left([g'(\mu)]^2 + g(\mu)g''(\mu)\right)^2 \text{var}\left[(z_i-\mu)^2\right]$$

$$+ \frac{1}{16}[g''(\mu)]^4 \text{var}\left[(z_i-\mu)^4\right] + [g'(\mu)]^2[g''(\mu)]^2 \text{var}\left[(z_i-\mu)^3\right]$$

$$+ 4g(\mu)[g'(\mu)]^2 g''(\mu) E\left[(z_i-\mu)^4\right]$$

$$+ \frac{1}{2}[g''(\mu)]^2 \left([g'(\mu)]^2 + g(\mu)g''(\mu)\right)\left(E\left[(z_i-\mu)^6\right] - \sigma^2 E\left[(z_i-\mu)^4\right]\right), \quad (3.13)$$

and

$$\text{var}\left[H(z_i)\right] \approx [H'(\mu)]^2 \sigma^2 + \frac{1}{4}[H''(\mu)]^2 \text{var}\left[(z_i-\mu)^2\right]. \quad (3.14)$$

We calculate the difference between $\text{var}\left[g^2(z_i)\right]$ and $\text{var}\left[H(z_i)\right]$.

$$\text{var}\left[g^2(z_i)\right] - \text{var}\left[H(z_i)\right] \approx \left(4g^2(\mu)[g'(\mu)]^2 - [H'(\mu)]^2\right)\sigma^2$$

$$+ \left[\left([g'(\mu)]^2 + g(\mu)g''(\mu)\right)^2 - \frac{1}{4}[H''(\mu)]^2\right] \text{var}\left[(z_i-\mu)^2\right]$$

$$+ \frac{1}{16}[g''(\mu)]^4 \text{var}\left[(z_i-\mu)^4\right] + [g'(\mu)]^2[g''(\mu)]^2 \text{var}\left[(z_i-\mu)^3\right]$$

$$+ 4g(\mu)[g'(\mu)]^2 g''(\mu) E\left[(z_i-\mu)^4\right]$$

$$+ \frac{1}{2}[g''(\mu)]^2 \left([g'(\mu)]^2 + g(\mu)g''(\mu)\right)\left(E\left[(z_i-\mu)^6\right] - \sigma^2 E\left[(z_i-\mu)^4\right]\right). \quad (3.15)$$

We notice that the third and forth items in (3.15) are automatically non-negative. Based on (3.15), we have two different situations:

I.

I.(i) $4g^2(\mu)[g'(\mu)]^2 - [H'(\mu)]^2 \geq 0$, so the first item in (3.15) is non-negative;

I.(ii) $\left([g'(\mu)]^2 + g(\mu)g''(\mu)\right)^2 - \frac{1}{4}[H''(\mu)]^2 \geq 0$, so the second item in (3.15) is non-negative;

I.(iii) $g(\mu)g''(\mu) \geq 0$, so the fifth item in (3.15) is non-negative.

I.(iv) $\left([g'(\mu)]^2 + g(\mu)g''(\mu)\right)$ and $E\left[(z_i-\mu)^6\right] - \sigma^2 E\left[(z_i-\mu)^4\right]$ are of the same sign so the last item in (3.15) is non-negative.

II.

The items in (3.15) are not all non-negative. Some of them are non-positive while the rest items are non-negative.

In situation I, we have $\text{var}[g^2(z_i)] - \text{var}[H(z_i)] \geq 0$ with certainty given the validity of the Taylor expansion, in which case (1.6) is a better choice than (1.5) to approximate FIN; in situation II, we can have a non-positive or non-negative value for $\text{var}[g^2(z_i)] - \text{var}[H(z_i)]$. The sign of $\text{var}[g^2(z_i)] - \text{var}[H(z_i)]$ cannot be found by checking the four conditions mentioned in situation I. Plugging in individual quantities for further calculation is necessary to figure this out.

We have found the sufficient conditions for $\text{var}[g^2(z_i)] - \text{var}[H(z_i)]$ to be non-negative, i.e. situation I. However, notice that our analysis does not provide necessary conditions for $\text{var}[g^2(z_i)] - \text{var}[H(z_i)]$ to be non-positive or non-negative. If we know that $\text{var}[g^2(z_i)] - \text{var}[H(z_i)]$ is, say, non-negative, we cannot conclude that all the items in (3.15) are all non-negative because the operations among them can give a non-negative value even if some of them are non-positive. As mentioned, further analysis is needed to explore this. In fact, finding the necessary conditions for $\text{var}[g^2(z_i)] - \text{var}[H(z_i)]$ to be non-positive or non-negative is a topic of interest for future study.

**Scenario B. Independent and non-identically distributed (i.n.i.d.) samples**

In the more general case where we have i.n.i.d. samples, the $z_i$ in $\mathbf{Z}_n = [z_1, z_2, ..., z_n]^T$ have density functions $p_i(\cdot|\theta)$, as well as mean $\mu_i$ and variance $\sigma_i^2$, $i = 1, 2, ..., n$. In order to be precise, we now include the subscript $i$ in $p_i(\cdot|\theta)$, $g_i^2(\cdot)$ and $H_i(\cdot)$ because the individual $p_i(\cdot|\theta)$, $g_i^2(\cdot)$ and $H_i(\cdot)$ are not necessarily identical.

According to the CLT, the following facts are valid:

$$\sqrt{n}\left(\bar{G}_n(\theta) - F(\theta)/n\right) \xrightarrow{dist} N(0, V_g), \tag{3.16}$$

$$\sqrt{n}\left(\bar{H}_n(\theta) - F(\theta)/n\right) \xrightarrow{dist} N(0, V_H), \tag{3.17}$$

where $V_g = \lim_{n\to\infty} n^{-1} \sum_{i=1}^{n} \text{var}[g_i^2(z_i)]$ and $V_H = \lim_{n\to\infty} n^{-1} \sum_{i=1}^{n} \text{var}[H_i(z_i)]$.

We calculate $\sum_{i=1}^{n} \text{var}[g_i^2(z_i)]$ and $\sum_{i=1}^{n} \text{var}[H_i(z_i)]$ and consider the value of $\sum_{i=1}^{n} \text{var}\left[g_i^2(z_i)\right] - \sum_{i=1}^{n} \text{var}\left[H_i(z_i)\right]$ in order to find out which one of (1.5) and (1.6) performs better in the i.n.i.d. samples scenario.

Define $\mu_i = E(z_i)$. For $i = 1, 2, \ldots, n$, let

$$g_i'(\mu_i) = \left.\frac{\partial g_i(z_i)}{\partial z_i}\right|_{z_i=\mu_i} \text{ and } g_i''(\mu_i) = \left.\frac{\partial^2 g_i(z_i)}{\partial z_i^2}\right|_{z_i=\mu_i};$$

$$H_i'(\mu_i) = \left.\frac{\partial H_i(z_i)}{\partial z_i}\right|_{z_i=\mu_i} \text{ and } H_i''(\mu_i) = \left.\frac{\partial^2 H_i(z_i)}{\partial z_i^2}\right|_{z_i=\mu_i}.$$

Then we have

$$\sum_{i=1}^{n} \text{var}\left[g_i^2(z_i)\right] \approx 4\sum_{i=1}^{n} g_i^2(\mu_i)\left[g_i'(\mu_i)\right]^2 \sigma_i^2$$

$$+ \sum_{i=1}^{n} \left([g_i'(\mu_i)]^2 + g_i(\mu_i)g_i''(\mu_i)\right)^2 \text{var}\left[(z_i - \mu_i)^2\right]$$

$$+ \frac{1}{16}\sum_{i=1}^{n} [g_i''(\mu_i)]^4 \text{var}\left[(z_i - \mu_i)^4\right] + \sum_{i=1}^{n} [g_i'(\mu_i)]^2 [g_i''(\mu_i)]^2 \text{var}\left[(z_i - \mu_i)^3\right]$$

$$+ 4\sum_{i=1}^{n} g_i(\mu_i)g_i'(\mu_i)\left([g_i'(\mu_i)]^2 + g_i(\mu_i)g_i''(\mu_i)\right) E\left[(z_i - \mu_i)^3\right]$$

$$+ \sum_{i=1}^{n} g_i(\mu_i)g_i'(\mu_i)[g_i''(\mu_i)]^2 E\left[(z_i - \mu_i)^5\right] + 4\sum_{i=1}^{n} g_i(\mu_i)[g_i'(\mu_i)]^2 g_i''(\mu_i) E\left[(z_i - \mu_i)^4\right]$$

$$+ \frac{1}{2}\sum_{i=1}^{n} [g_i''(\mu_i)]^2 \left([g_i'(\mu_i)]^2 + g_i(\mu_i)g_i''(\mu_i)\right)\left(E\left[(z_i - \mu_i)^6\right] - \sigma_i^2 E\left[(z_i - \mu_i)^4\right]\right)$$

$$+ 2\sum_{i=1}^{n} g_i'(\mu_i)g_i''(\mu_i)\left([g_i'(\mu_i)]^2 + g_i(\mu_i)g_i''(\mu_i)\right)\left(E\left[(z_i - \mu_i)^5\right] - \sigma_i^2 E\left[(z_i - \mu_i)^3\right]\right)$$

$$+ \frac{1}{2}\sum_{i=1}^{n} g_i'(\mu_i)[g_i''(\mu_i)]^3 \left(E\left[(z_i - \mu_i)^7\right] - E\left[(z_i - \mu_i)^4\right] E\left[(z_i - \mu_i)^3\right]\right), \tag{3.18}$$

and

$$\sum_{i=1}^{n} \text{var}[H_i(z_i)] \approx \sum_{i=1}^{n} \left[H_i'(\mu_i)\right]^2 \sigma_i^2 + \frac{1}{4}\sum_{i=1}^{n} \left[H_i''(\mu_i)\right]^2 \text{var}\left[(z_i - \mu_i)^2\right]$$

$$+ \sum_{i=1}^{n} H_i'(\mu_i)H_i''(\mu_i) E\left\{[z_i - \mu_i]^3\right\}. \tag{3.19}$$

Still, we consider symmetric density functions. Thus simplifying (3.16) and (3.17) to

$$\sum_{i=1}^{n} \mathrm{var}\left[g_i^2(z_i)\right] \approx 4\sum_{i=1}^{n} g_i^2(\mu_i)\left[g_i'(\mu_i)\right]^2 \sigma_i^2$$

$$+\sum_{i=1}^{n}\left([g_i'(\mu_i)]^2 + g_i(\mu_i)g_i''(\mu_i)\right)^2 \mathrm{var}\left[(z_i-\mu_i)^2\right]$$

$$+\frac{1}{16}\sum_{i=1}^{n}[g_i''(\mu_i)]^4 \mathrm{var}\left[(z_i-\mu_i)^4\right] + \sum_{i=1}^{n}[g_i'(\mu_i)]^2[g_i''(\mu_i)]^2 \mathrm{var}\left[(z_i-\mu_i)^3\right]$$

$$+4\sum_{i=1}^{n} g_i(\mu_i)[g_i'(\mu_i)]^2 g_i''(\mu_i) E\left[(z_i-\mu_i)^4\right]$$

$$+\frac{1}{2}\sum_{i=1}^{n}[g_i''(\mu_i)]^2\left([g_i'(\mu_i)]^2 + g_i(\mu_i)g_i''(\mu_i)\right)\left(E\left[(z_i-\mu_i)^6\right] - \sigma_i^2 E\left[(z_i-\mu_i)^4\right]\right), \quad (3.20)$$

and

$$\sum_{i=1}^{n} \mathrm{var}[H_i(z_i)] \approx \sum_{i=1}^{n}\left[H_i'(\mu_i)\right]^2 \sigma_i^2 + \frac{1}{4}\sum_{i=1}^{n}\left[H_i''(\mu_i)\right]^2 \mathrm{var}\left[(z_i-\mu_i)^2\right]. \quad (3.21)$$

Next, we consider the difference between $\sum_{i=1}^{n}\mathrm{var}\left[g_i^2(z_i)\right]$ and $\sum_{i=1}^{n}\mathrm{var}[H_i(z_i)]$.

$$\sum_{i=1}^{n}\mathrm{var}\left[g_i^2(z_i)\right] - \sum_{i=1}^{n}\mathrm{var}[H_i(z_i)]$$

$$\approx \sum_{i=1}^{n}\left(4g_i^2(\mu_i)\left[g_i'(\mu_i)\right]^2 - \left[H_i'(\mu_i)\right]^2\right)\sigma_i^2$$

$$+\sum_{i=1}^{n}\left[\left([g_i'(\mu_i)]^2 + g_i(\mu_i)g_i''(\mu_i)\right)^2 - \frac{1}{4}\left[H_i''(\mu_i)\right]^2\right]\mathrm{var}\left[(z_i-\mu_i)^2\right]$$

$$+\frac{1}{16}\sum_{i=1}^{n}[g_i''(\mu_i)]^4 \mathrm{var}\left[(z_i-\mu_i)^4\right] + \sum_{i=1}^{n}[g_i'(\mu_i)]^2[g_i''(\mu_i)]^2 \mathrm{var}\left[(z_i-\mu_i)^3\right]$$

$$+4\sum_{i=1}^{n} g_i(\mu_i)[g_i'(\mu_i)]^2 g_i''(\mu_i) E\left[(z_i-\mu_i)^4\right]$$

$$+\frac{1}{2}\sum_{i=1}^{n}[g_i''(\mu_i)]^2\left([g_i'(\mu_i)]^2 + g_i(\mu_i)g_i''(\mu_i)\right)\left(E\left[(z_i-\mu_i)^6\right] - \sigma_i^2 E\left[(z_i-\mu_i)^4\right]\right). \quad (3.22)$$

Notice that the third and forth items in the above result is automatically non-positive. As with the i.i.d. case, we have two situations based on (3.22):

I′.

I′(i)　For all $i = 1, 2, …, n$, $4g_i^2(\mu_i)\left[g_i'(\mu_i)\right]^2 - \left[H_i'(\mu_i)\right]^2 \geq 0$ so that the first item in (3.22) is non-negative;

I′(ii)　For all $i = 1, 2, …, n$, $\left(\left[g_i'(\mu_i)\right]^2 + g_i(\mu_i)g_i''(\mu_i)\right)^2 - \frac{1}{4}\left[H_i''(\mu_i)\right]^2 \geq 0$ so that the second item in

(3.22) is non-negative;

I′(iii) For all $i = 1, 2, \ldots, n$, $g(\mu)g''(\mu) \geq 0$ so that the fifth item in (3.22) is non-negative.

I′.(iv) $\left([g_i'(\mu_i)]^2 + g_i(\mu_i)g_i''(\mu_i)\right)$ and $E\left[(z_i - \mu_i)^6\right] - \sigma_i^2 E\left[(z_i - \mu_i)^4\right]$ are of the same sign so the last item in (3.22) is non-negative.

II′.

The items in (3.22) are not all non-negative. Some of them are non-negative while the rest items are non-positive.

In situation I′, we have $\sum_{i=1}^{n} \text{var}\left[g_i^2(z_i)\right] - \sum_{i=1}^{n} \text{var}\left[H_i(z_i)\right] \geq 0$ with certainty, in which case (1.6) is a better choice than (1.5) to approximate FIN; in situation II′, we can have either a non-positive or non-negative value for $\sum_{i=1}^{n} \text{var}\left[g_i^2(z_i)\right] - \sum_{i=1}^{n} \text{var}\left[H_i(z_i)\right]$, which cannot be found by checking the four conditions mentioned in I′. Further calculation is necessary to lead to a result indicating the sign of $\sum_{i=1}^{n} \text{var}\left[g_i^2(z_i)\right] - \sum_{i=1}^{n} \text{var}\left[H_i(z_i)\right]$.

So far, the analysis has led us to the sufficient conditions for $\sum_{i=1}^{n} \text{var}\left[g_i^2(z_i)\right] - \sum_{i=1}^{n} \text{var}\left[H_i(z_i)\right]$ to be non-negative, i.e. situation I′. Nonetheless, our analysis does not provide the necessary conditions. For instance, if we know that $\sum_{i=1}^{n} \text{var}\left[g_i^2(z_i)\right] - \sum_{i=1}^{n} \text{var}\left[H_i(z_i)\right]$ is non-negative, we cannot conclude that the items in (3.22) are all non-negative because the operations among them can give a non-negative value even if some of them are non-positive. Further analysis is needed to address this issue. One of the proposals for future study is to find out the necessary conditions for $\sum_{i=1}^{n} \text{var}\left[g_i^2(z_i)\right] - \sum_{i=1}^{n} \text{var}\left[H_i(z_i)\right]$ to be non-positive or non-negative. Hopefully this course will be figured out in future work.

Therefore, given the validity of the Taylor expansion-based analysis, we have reached certain results for both i.i.d. sample cases and i.n.i.d. sample cases, though they do not appear succinct enough at the first glance and thus raising concerns about their applicability. It turns out that they fit in well in practice. In effect, for the many density functions that we confront in practice, especially the exponential family, the above results can be substantially simplified and we are able to apply them to check which one of (1.5) and (1.6) does a better job in estimating the FIN. Illustrations are provided in the section of numerical studies.

## 4. Numerical Studies

We have done theoretical analysis. Next we would like to present some case studies as demonstrations.

**Example 1—Normal distribution** $N(\mu,\sigma^2)$ **with** $\theta = \mu$

We look at a trivial case first. We follow the notations used in the theoretical analysis. Consider an i.i.d. sample set where $\mathbf{Z}_n = [z_1, z_2, ..., z_n]$ is normally distributed with mean $\mu$ and variance $\sigma^2$. The unknown parameter is $\theta = \mu$. The density function is $p(z|\theta) = (2\pi\sigma^2)^{-1/2} \exp\{-(z-\mu)^2/2\sigma^2\}$. Hence the log-density function is

$$\log p(z|\mu) = -\frac{1}{2}\log 2\pi - \frac{1}{2}\log \sigma^2 - \frac{1}{2\sigma^2}(z-\mu)^2. \tag{4.1}$$

The gradient and the Hessian for an arbitrary random variable in the data set are given by:

$$g(z|\mu) = \frac{\partial \log p(z|\mu)}{\partial \mu} = \frac{(z-\mu)}{\sigma^2}, \tag{4.2}$$

$$H(z|\mu) = \frac{\partial^2 \log p(z|\mu)}{\partial \mu^2} = -\frac{1}{\sigma^2}. \tag{4.3}$$

We calculate $g(\mu^*)$, $g'(\mu^*)$, $g''(\mu^*)$, $H(\mu^*)$, $H'(\mu^*)$ and $H''(\mu^*)$, where $\mu^*$ denotes the true value of the unknown parameter. Then

$$g(\mu^*) = \frac{1}{\sigma^2}(\mu^* - \mu), \; g'(\mu^*) = \frac{\partial g(z)}{\partial z}\bigg|_{z=\mu^*} = \frac{1}{\sigma^2} \text{ and } g''(\mu^*) = \frac{\partial^2 g(z)}{\partial z^2}\bigg|_{z=\mu^*} = 0; \tag{4.4}$$

$$H(\mu^*) = -\frac{1}{\sigma^2}, H'(\mu^*) = \frac{\partial H(z)}{\partial z}\bigg|_{z=\mu^*} = 0 \text{ and } H''(\mu^*) = \frac{\partial^2 H(z)}{\partial z^2}\bigg|_{z=\mu^*} = 0. \tag{4.5}$$

Using the above facts, we check which situations among I, II and III is a fit for $N(\mu,\sigma^2)$.

First, $4g^2(\mu^*)[g'(\mu^*)]^2 - [H'(\mu^*)]^2 = 4\frac{1}{\sigma^4}\left[\frac{1}{\sigma^2}(\mu^* - \mu)\right]^2 \geq 0$;

Next, $\left([g'(\mu^*)]^2 + g(\mu^*)g''(\mu^*)\right)^2 - \frac{1}{4}[H''(\mu^*)]^2 = \frac{1}{\sigma^8} \geq 0$;

- 14 -

This shows that situation I in the i.i.d. scenario is satisfied, which means that $\text{var}\left[g^2(z)\right] - \text{var}\left[H(z)\right] > 0$, implying the fact that for normally distributed i.i.d. samples with mean $\mu$ and variance $\sigma^2$, the average of second-order derivatives in (1.6) should outperform the average of squared derivatives in (1.5) as an approximation for FIN.

Nonetheless, this conclusion should have been made initially. We know that for $N(\mu, \sigma^2)$, the second-order derivative of the log-density function equals to a constant $-(\sigma^2)^{-1}$. The variance $\text{var}\left[H(z)\right] = \text{var}[(\sigma^2)^{-1}] = 0$. $\text{var}\left[H(z)\right]$ is automatically no larger than $\text{var}\left[g^2(z)\right]$, indicating that we should choose (1.6) over (1.5) to be the estimator of FIN.

To check if this is true, we use Matlab to generate normal random numbers and calculate (1.5) and (1.6) to see whether (1.6) has a smaller variance than (1.5). Consider three cases where we have (1) $\mu = 1$ and $\sigma^2 = 0.1^2$ (2) $\mu = 0$ and $\sigma^2 = 1$ (3) $\mu = 5$ and $\sigma^2 = 10^2$.

For a Monte Carlo simulation with sample size of $n = 100$ and $N = 200$ replicates, the results are as follows:

**Table 1.a** Sample variances of (1.5) and (1.6) as well as the difference between them, when the unknown parameter $\theta = \mu$

| | $\mu = 1$ and $\sigma^2 = 0.1^2$ | $\mu = 0$ and $\sigma^2 = 1$ | $\mu = 5$ and $\sigma^2 = 10^2$ |
|---|---|---|---|
| $\text{var}^{(e)}\left[g^2(z)\right]$ | 87.3516 | 95.4156 | 95.2212 |
| $\text{var}^{(e)}\left[H(z)\right]$ | 0 | 0 | $4.8635 \times 10^{-35}$ |
| $\text{var}^{(e)}\left[g^2(z)\right] - \text{var}^{(e)}\left[H(z)\right]$ | 87.3516 | 95.4156 | 95.2212 |

Here $\text{var}^{(e)}(X)$ stands for the sample variance of the random variable $X$, which is defined as $\text{var}^{(e)}(X) = (n-1)^{-1} \sum_{i=1}^{n} (x_i - \bar{x})^2$, $\bar{x} = n^{-1} \sum_{i=1}^{n} x_i$ with $x_i$ denoting the $i$th sample value, $i = 1, 2, ..., n$. It is an unbiased estimator of the theoretical variance and its value is analytically calculated in order to illustrate the magnitude of the true asymptotic variance. For the following context, we will use the same notation for the same meaning and purpose.

We do a hypothesis test to see whether $\text{var}^{(e)}\left[g^2(z)\right] - \text{var}^{(e)}\left[H(z)\right]$ is significantly larger than 0. Run a $t$-test for $m = 50$ replicates, we have:

**Table 1.b** Hypothesis testing results for $\theta = \mu$ using a one-sided $t$-test with 50 replicates

|  | $\mu = 1$ and $\sigma^2 = 0.1^2$ | $\mu = 0$ and $\sigma^2 = 1$ | $\mu = 5$ and $\sigma^2 = 10^2$ |
|---|---|---|---|
| $P$-value | $3.0776 \times 10^{-25}$ | $8.0637 \times 10^{-33}$ | $2.0196 \times 10^{-19}$ |

As shown, the outcomes for these three cases all coincide with the conclusion of the theoretical analysis. As $\sigma^2$ gets larger, $\text{var}^{(e)}\left[g^2(z)\right]$ and $\text{var}^{(e)}\left[H(z)\right]$ tend to increase, so does $\text{var}^{(e)}\left[g^2(z)\right] - \text{var}^{(e)}\left[H(z)\right]$.

**Example 2—Normal distribution** $N(\mu, \sigma^2)$ **with** $\theta = \sigma^2$

We still consider an i.i.d. sample set where $\mathbf{Z}_n = [z_1, z_2, ..., z_n]$ is normally distributed with mean $\mu$ and variance $\sigma^2$, but this time the unknown parameter is $\theta = \sigma^2$ instead of $\theta = \mu$ and we assume that $\mu$ is known. The density function is the same with the one in Example 1, so is the log-density function. Yet the gradient and Hessian change to

$$g(\sigma^2 \mid z) = \frac{\partial \log p(z \mid \sigma^2)}{\partial \sigma^2} = -\frac{1}{2\sigma^2} + \frac{(z-\mu)^2}{2\sigma^4}, \tag{4.6}$$

$$H(\sigma^2 \mid z) = \frac{\partial^2 \log p(z \mid \sigma^2)}{\partial (\sigma^2)^2} = \frac{1}{2\sigma^4} - \frac{(z-\mu)^2}{\sigma^6}. \tag{4.7}$$

We calculate $g(\mu)$, $g'(\mu)$, $g''(\mu)$, $H(\mu)$, $H'(\mu)$ and $H''(\mu)$.

$$g(\mu) = -\frac{1}{2\sigma^2}, \; g'(\mu) = \frac{\partial g(z)}{\partial z}\bigg|_{z=\mu} = 0, \; g''(\mu) = \frac{\partial^2 g(z)}{\partial z^2}\bigg|_{z=\mu} = \frac{1}{\sigma^4}; \tag{4.8}$$

$$H(\mu) = \frac{1}{2\sigma^4}, \; H'(\mu) = \frac{\partial H(z)}{\partial z}\bigg|_{z=\mu} = 0, \; H''(\mu) = \frac{\partial^2 H(z)}{\partial z^2}\bigg|_{z=\mu} = -\frac{2}{\sigma^6}. \tag{4.9}$$

It is easily seen that this example does not fit in situation I, which means that we should plug in $g(\mu)$, $g'(\mu), g''(\mu), H(\mu), H'(\mu)$ and $H''(\mu)$ to get the approximated value for $\text{var}\left[g^2(z)\right] - \text{var}\left[H(z)\right]$.

$$\text{var}\left[g^2(z)\right] - \text{var}[H(z)]$$

$$\approx \left(\frac{1}{4\sigma^{12}} - \frac{1}{\sigma^8}\right) \cdot 2\sigma^4 + \frac{1}{16\sigma^{16}}\left(7!!\sigma^8 - (3\sigma^4)^2\right) + \frac{1}{2\sigma^8}\left(-\frac{1}{2\sigma^6}\right)\left(5!!\sigma^6 - 3\sigma^6\right) = \frac{3}{2\sigma^8} > 0, \quad (4.10)$$

where $a!! = a(a-2)(a-4)(a-6)\cdots \times 3 \times 1$ for some integer $a$.

Therefore, we assert that $\text{var}\left[g^2(z)\right] - \text{var}[H(z)] \geq 0$. This implies that for i.i.d. $N(\mu,\sigma^2)$ samples with $\theta = \sigma^2$, (1.6) outperforms (1.5), theoretically. Notice that this example is non-trivial because the Hessian matrix is no longer a constant (i.e. it is a function of $\theta$). We cannot tell which form between (1.5) and (1.6) is optimal without analysis.

Again, we use Matlab to generate normal random numbers and calculate (1.5) and (1.6) to see whether (1.6) has a smaller asymptotic variance than (1.5). Consider the three cases we used in Example 1 (1) $\mu = 1$ and $\sigma^2 = 0.1^2$ (2) $\mu = 0$ and $\sigma^2 = 1$ (3) $\mu = 5$ and $\sigma^2 = 10^2$.

For a program using a single run of $n = 100$ and replicates of $N = 200$, the results are as follows:

**Table 2.a** Sample variances of (1.5) and (1.6) as well as the ratio of the former variance to the latter variance when the unknown parameter $\theta = \sigma^2$

|  | $\mu = 1$ and $\sigma^2 = 0.1^2$ | $\mu = 0$ and $\sigma^2 = 1$ | $\mu = 5$ and $\sigma^2 = 10^2$ |
|---|---|---|---|
| $\text{var}^{(e)}\left[g^2(z)\right]$ | $3.1188 \times 10^8$ | 3.6371 | $3.1812 \times 10^{-8}$ |
| $\text{var}^{(e)}[H(z)]$ | $1.9226 \times 10^8$ | 2.0632 | $1.9380 \times 10^{-8}$ |
| $\text{var}^{(e)}\left[g^2(z)\right]/\text{var}^{(e)}[H(z)]$ | 1.6222 | 1.7628 | 1.6415 |

We use the ratio of the two variances instead of their difference in the above table. A ratio larger than 1 implies that $\text{var}^{(e)}\left[g^2(z)\right]$ is larger than $\text{var}^{(e)}[H(z)]$. In this case the numerical results show that (1.6) does outperform (1.5) in the sense that (1.6) has a smaller asymptotic variance than (1.5) does. One noteworthy phenomenon is that as the true $\sigma^2$ gets larger, the asymptotic variances of the two methods tend to get smaller, which is a little counterintuitive. However, as rarely as it is, this fact is justified in the sense that the unknown parameter we estimate is in the denominator so the magnitude of the variance is inversely related to its value.

As in the last case study, we do a *t*-test for $\text{var}^{(e)}\left[g^2(z)\right] - \text{var}^{(e)}\left[H(z)\right]$ with $m = 50$ replicates:

**Table 2.b** Hypothesis testing results for $\theta = \sigma^2$ using a one-sided t-test with 50 replicates

|  | $\mu = 1$ and $\sigma^2 = 0.1^2$ | $\mu = 0$ and $\sigma^2 = 1$ | $\mu = 5$ and $\sigma^2 = 10^2$ |
|---|---|---|---|
| *P-value* | $7.7846 \times 10^{-5}$ | $3.2146 \times 10^{-5}$ | $9.4938 \times 10^{-5}$ |

The results from our hypothesis test indicate that $\text{var}^{(e)}\left[g^2(z)\right] - \text{var}^{(e)}\left[H(z)\right]$ is significantly different from 0, i.e. larger than 0 in this example. Therefore, (1.6) indeed outperforms (1.5) in the sense that it is more accurate and gives a smaller variance for the estimated FIN. The prevalence of (1.6) as a method to approximate FIN is justified for this individual situation.

**Example 3—Signal-plus noise**

We also present a case study for i.n.i.d. scenario. Consider the signal-plus noise problem where we have i.n.i.d. data with $z_i \sim N(0, \sigma^2 + q_i)$, $q_i = 0.1 \times (i - 10 \times \lfloor i/10 \rfloor)$ where $\lfloor i/10 \rfloor$ denotes the integer part of $i/10$, $i = 1, 2, \ldots, n$. We would like to estimate the unknown parameter $\theta = \sigma^2$. For an arbitrary $i$,

$$p(z_i \mid \sigma^2) = \frac{1}{\sqrt{2\pi(\sigma^2 + q_i)}} \exp\left\{-\frac{z_i^2}{2(\sigma^2 + q_i)}\right\}. \tag{4.11}$$

Thus, the log-density function is

$$\log p(z_i \mid \sigma^2) = -\frac{1}{2}\log 2\pi - \frac{1}{2}\log(\sigma^2 + q_i) - \frac{z_i^2}{2(\sigma^2 + q_i)}. \tag{4.12}$$

Next we compute the gradient and the Hessian:

$$g(\sigma^2 \mid z_i) = \frac{\partial \log p(z_i \mid \sigma^2)}{\partial \sigma^2} = -\frac{1}{2(\sigma^2 + q_i)} + \frac{z_i^2}{2(\sigma^2 + q_i)^2}, \tag{4.13}$$

$$H(\sigma^2 \mid z_i) = \frac{\partial^2 \log p(z_i \mid \sigma^2)}{\partial (\sigma^2)^2} = \frac{1}{2(\sigma^2 + q_i)^2} - \frac{z_i^2}{(\sigma^2 + q_i)^3}. \tag{4.14}$$

We calculate $g(0)$, $g'(0)$, $g''(0)$, $H(0)$, $H'(0)$ and $H''(0)$.

$$g_i(0) = -\frac{1}{2(\sigma^2 + q_i)}, \; g_i'(0) = \left.\frac{z_i}{(\sigma^2 + q_i)^2}\right|_{z_i=0} = 0, \; g_i''(0) = \frac{1}{(\sigma^2 + q_i)^2}; \tag{4.15}$$

$$H_i(0) = \frac{1}{2(\sigma^2+q_i)^2}, H_i'(0) = -\frac{2z_i}{(\sigma^2+q_i)^3}\bigg|_{z_i=0} = 0, H_i''(0) = -\frac{2}{(\sigma^2+q_i)^3}. \tag{4.16}$$

Again, this example doesn't fit in situation I′, which means that we should plug in $g(\mu)$, $g'(\mu)$, $g''(\mu)$, $H(\mu)$, $H'(\mu)$ and $H''(\mu)$ to get the approximated value for $\text{var}\left[g_i^2(z_i)\right] - \text{var}\left[H_i(z_i)\right]$.

$$\begin{aligned}
&\text{var}\left[g_i^2(z_i)\right] - \text{var}\left[H_i(z_i)\right] \\
&\approx \left(\frac{1}{4(\sigma^2+q_i)^6} - \frac{1}{(\sigma^2+q_i)^6}\right) \cdot 2(\sigma^2+q_i)^2 + \frac{1}{16(\sigma^2+q_i)^8}\left(7!!(\sigma^2+q_i)^4 - (3(\sigma^2+q_i)^2)^2\right) \\
&\quad + \frac{1}{2(\sigma^2+q_i)^4}\left(-\frac{1}{2(\sigma^2+q_i)^3}\right)\left(5!!(\sigma^2+q_i)^3 - 3(\sigma^2+q_i)^3\right) \\
&= \frac{3}{2(\sigma^2+q_i)^4} > 0.
\end{aligned} \tag{4.17}$$

Note that we know whether $\sum_{i=1}^n \text{var}\left[g_i^2(z_i)\right] - \sum_{i=1}^n \text{var}\left[H_i(z_i)\right]$ is non-positive or non-negative by finding out whether $\text{var}\left[g_i^2(z_i)\right] - \text{var}\left[H_i(z_i)\right]$ is non-positive or non-negative. Given the above analysis, we assert that $\text{var}\left[g_i^2(z_i)\right] - \text{var}\left[H_i(z_i)\right] \geq 0$. Theoretically (1.6) should outperform (1.5) in the sense that it has a smaller asymptotic variance. We check this assertion by implementing the example in Matlab.

Using three numerical cases where (1) $\sigma^2 = 0.1^2$ (2) $\sigma^2 = 1$ (3) $\sigma^2 = 10^2$, with $n = 100$ and $N = 200$, we get the following results:

**Table 3.a** Sample variances of (1.5) and (1.6) as well as the ratio of the former variance to the latter variance when the unknown parameter $\theta = \sigma^2$.

|  | $\sigma^2 = 0.1^2$ | $\sigma^2 = 1$ | $\sigma^2 = 10^2$ |
|---|---|---|---|
| $\text{var}^{(e)}\left[g_i^2(z_i)\right]$ | $6.1060 \times 10^7$ | 1.0349 | $3.5680 \times 10^{-8}$ |
| $\text{var}^{(e)}\left[H_i(z_i)\right]$ | $2.5194 \times 10^7$ | 0.6584 | $2.0023 \times 10^{-8}$ |
| $\text{var}^{(e)}\left[g_i^2(z_i)\right]/\text{var}^{(e)}\left[H_i(z_i)\right]$ | 2.4236 | 1.5718 | 1.7820 |

One noteworthy phenomenon is the recurrence of the situation confronted in Example 2 — both variances decrease as $\sigma^2$ grows, which is uncommon but not unacceptable. Clearly, the numerical results coincide with our assertion that (1.6) leads to a smaller asymptotic variance. In order to validate the results, we do a *t*-test for $\text{var}\left[g_i^2(z_i)\right] - \text{var}\left[H_i(z_i)\right]$:

**Table 3.b** Hypothesis testing results for $\theta=\sigma^2$ using a one-sided t-test with 50 replicates

|  | $\sigma^2 = 0.1^2$ | $\sigma^2 = 1$ | $\sigma^2 = 10^2$ |
|---|---|---|---|
| *P-value* | 0.0269 | 0.0025 | $2.3863 \times 10^{-7}$ |

As shown in Table 3.b, the *t*-test demonstrates that the asymptotic variances of (1.5) and (1.6) in the i.n.i.d. scenario are significantly different, i.e. bigger than 0. This fact supports the utilization of (1.6) as the way to estimate FIN instead of (1.5) for this application.

# 5. Conclusions and Future Work

**5.1 Conclusion**

So far we have looked at the comparison of accuracy for the two approximation methods in the scalar case and mainly focused our analysis on the family of symmetric density functions. We have considered i.i.d data and i.n.i.d data and get some results for both scenarios.

The example for i.n.i.d data, single-plus noise for the scalar case, has a variety of utilizations in practice. According to our theoretical results, when dealing with this sort of problem, the Hessian method is a better choice because it gives a smaller asymptotic variance than the product-of-gradient method does. The estimated difference between their asymptotic variances is $3 \cdot (200)^{-1} \sum_{i=1}^{100} (\sigma^2 + q_i)^{-4}$. The numerical study confirms our analysis. This is also the case where the density function is normal with the unknown parameter being estimated is $\sigma^2$. The difference between these two variances is approximately $3 \cdot (2\sigma^8)^{-1}$ according to our theory.

Some interesting phenomena indeed appear during the research procedure. A salient one arises in the utilization of the Taylor expansion. Note that we estimate $\text{var}\left[g^2(z_i)\right]$ (or $\text{var}\left[g_i^2(z_i)\right]$) with the second-order Taylor expansion by first expanding $g(z_i)$ (or $g_i(z_i)$) and then squaring it and proceed to the rest of the calculation. We don't expand $g^2(z_i)$ (or $g_i^2(z_i)$) directly. This is a tricky part of the study because if we do this in the latter way, we are not able to come up with a valid analysis for the magnitude of the two asymptotic variances. Rather, we find that the magnitude of $\text{var}\left[g^2(z_i)\right]$ (or $\text{var}\left[g_i^2(z_i)\right]$) would be substantially underestimated. Some non-negative items would be neglected because of the incorrect use of the Taylor expansion, thus resulting in an approximation much smaller than the true value.

**5.2 Future Work**

In this report, we mainly consider the situation when the explicit form of the density function is accessible, but the true value of Fisher Information is not readily available so we have to estimate it. We compare the accuracy of the methods we use to achieve this. The magnitudes of the asymptotic variances of two approximation methods are contrasted and a conclusion is made that the accuracy is case dependent. A couple of conditions are necessary to be looked at to determine which method is to be used. Due to the limitation of time and length of the report, we have only considered the case where the unknown parameter $\theta$ is scalar. Two scenarios are looked at, one of which is when we have i.i.d samples and the other one is when we have i.n.i.d samples. Examples for both scenarios are provided in the numerical studies, which illustrate our conclusion by numerical results.

Nevertheless, this report has its limitations. It fails to consider the multidimensional case where $\theta$ is a vector. In reality, a vector $\theta$ of $p$ dimension where $p$ is no less than 2 is often the case we confront. Truly this brings complexity to our analysis. Also, our study mainly concentrates on the symmetric density functions. A broader scope of case study is desired, which can be a potential direction for future study. Furthermore, our theoretical analysis merely gives the sufficient conditions to determine the values for the key quantities $\text{var}\left[g^2(z_i)\right] - \text{var}\left[H(z_i)\right]$ and $\sum_{i=1}^{n}\text{var}\left[g_i^2(z_i)\right] - \sum_{i=1}^{n}\text{var}\left[H_i(z_i)\right]$. Finding the necessary conditions to decide the magnitudes of these two key quantities is a promising research direction.

Obviously, there is more work to do for our research in future. Three topics to work on in future are listed below, which are representative but not comprehensive:

1. Apply the sufficient conditions we get in this paper to density functions other than the exponential family and see whether the numerical results match the theoretical analysis;
2. Find out the necessary conditions for the two key quantities and $\sum_{i=1}^{n}\text{var}\left[g_i^2(z_i)\right] - \sum_{i=1}^{n}\text{var}\left[H_i(z_i)\right]$ to be non-positive or non-negative;
3. Generalize the study to multidimensional situation where $\theta$ is a vector.

# 6. Appendix

In this section, we present the detailed calculation of $\text{var}\left[g^2(z_i)\right]$, $\text{var}[H(z_i)]$, $\text{var}\left[g_i^2(z_i)\right]$, $\text{var}[H_i(z_i)]$, $\sum_{i=1}^{n}\text{var}\left[g_i^2(z_i)\right]$ and $\sum_{i=1}^{n}\text{var}[H_i(z_i)]$. As mentioned before, the calculation of the actual values of $\text{var}\left[g^2(z_i)\right]$, $\text{var}[H(z_i)]$, $\text{var}\left[g_i^2(z_i)\right]$, $\text{var}[H_i(z_i)]$, $\sum_{i=1}^{n}\text{var}\left[g_i^2(z_i)\right]$ and $\sum_{i=1}^{n}\text{var}[H_i(z_i)]$ is demanding. We utilize the second-order Taylor expansion to give approximations to these values.

For i.i.d. samples, define $E(z_i)=\mu$. In order to derive (4.11) and (4.12), we do second-order Taylor expansion to $g(z_i)$ at $z_i=\mu$, $i=1, 2, \ldots, n$. Using the notation before, we have

$$g(z_i) \approx g(\mu) + g'(\mu)(z_i-\mu) + \frac{1}{2}g''(\mu)(z_i-\mu)^2, \tag{6.1}$$

Square it, we get

$$g^2(z_i) \approx \left(g(\mu) + g'(\mu)(z_i-\mu) + \frac{1}{2}g''(\mu)(z_i-\mu)^2\right)^2$$

$$= g^2(\mu) + [g'(\mu)]^2(z_i-\mu)^2 + \frac{1}{4}[g''(\mu)]^2(z_i-\mu)^4$$

$$+ 2g(\mu)g'(\mu)(z_i-\mu) + g(\mu)g''(\mu)(z_i-\mu)^2 + g'(\mu)g''(\mu)(z_i-\mu)^3. \tag{6.2}$$

Take expectation on both sides,

$$E[g^2(z_i)] \approx g^2(\mu) + \left([g'(\mu)]^2 + g(\mu)g''(\mu)\right)\sigma^2 + \frac{1}{4}[g''(\mu)]^2 E[(z_i-\mu)^4] + g'(\mu)g''(\mu)E[(z_i-\mu)^3]. \tag{6.3}$$

Therefore,

$$\text{var}[g^2(z_i)] = E\left[\left(g^2(z_i) - E[g^2(z_i)]\right)^2\right]$$

$$\approx E\Bigg[\Bigg([g'(\mu)]^2(z_i-\mu)^2 + \frac{1}{4}[g''(\mu)]^2(z_i-\mu)^4$$

$$+ 2g(\mu)g'(\mu)(z_i-\mu) + g(\mu)g''(\mu)(z_i-\mu)^2 + g'(\mu)g''(\mu)(z_i-\mu)^3$$

$$- \left([g'(\mu)]^2 + g(\mu)g''(\mu)\right)\sigma^2 - \frac{1}{4}[g''(\mu)]^2 E[(z_i-\mu)^4] - g'(\mu)g''(\mu)E[(z_i-\mu)^3]\Bigg)^2\Bigg]$$

$$= E\Bigg[\Bigg(2g(\mu)g'(\mu)(z_i-\mu) + \left([g'(\mu)]^2 + g(\mu)g''(\mu)\right)\left((z_i-\mu)^2 - \sigma^2\right)$$

$$+\frac{1}{4}[g''(\mu)]^2\left((z_i-\mu)^4 - E[(z_i-\mu)^4]\right) + g'(\mu)g''(\mu)\left((z_i-\mu)^3 - E[(z_i-\mu)^3]\right)\Big)^2\Bigg]$$

$$= E\Bigg[4g^2(\mu)[g'(\mu)]^2(z_i-\mu)^2 + \left([g'(\mu)]^2 + g(\mu)g''(\mu)\right)^2\left((z_i-\mu)^2 - \sigma^2\right)^2$$

$$+\frac{1}{16}[g''(\mu)]^4\left((z_i-\mu)^4 - E[(z_i-\mu)^4]\right)^2 + [g'(\mu)]^2[g''(\mu)]^2\left((z_i-\mu)^3 - E[(z_i-\mu)^3]\right)^2$$

$$+4g(\mu)g'(\mu)\left([g'(\mu)]^2 + g(\mu)g''(\mu)\right)\left((z_i-\mu)^3 - \sigma^2(z_i-\mu)\right)$$

$$+g(\mu)g'(\mu)[g''(\mu)]^2\left((z_i-\mu)^5 - (z_i-\mu)E[(z_i-\mu)^4]\right)$$

$$+4g(\mu)[g'(\mu)]^2 g''(\mu)\left((z_i-\mu)^4 - (z_i-\mu)E[(z_i-\mu)^3]\right)$$

$$+\frac{1}{2}[g''(\mu)]^2\left([g'(\mu)]^2 + g(\mu)g''(\mu)\right)\underbrace{\left((z_i-\mu)^2 - \sigma^2\right)\left((z_i-\mu)^4 - E[(z_i-\mu)^4]\right)}_{(a)}$$

$$+2g'(\mu)g''(\mu)\left([g'(\mu)]^2 + g(\mu)g''(\mu)\right)\underbrace{\left((z_i-\mu)^2 - \sigma^2\right)\left((z_i-\mu)^3 - E[(z_i-\mu)^3]\right)}_{(b)}$$

$$+\frac{1}{2}g'(\mu)[g''(\mu)]^3\underbrace{\left((z_i-\mu)^3 - E[(z_i-\mu)^3]\right)\left((z_i-\mu)^4 - E[(z_i-\mu)^4]\right)}_{(c)}\Bigg]$$

$$= 4g^2(\mu)[g'(\mu)]^2\sigma^2 + \left([g'(\mu)]^2 + g(\mu)g''(\mu)\right)^2 \mathrm{var}\left[(z_i-\mu)^2\right]$$

$$+\frac{1}{16}[g''(\mu)]^4 \mathrm{var}\left[(z_i-\mu)^4\right] + [g'(\mu)]^2[g''(\mu)]^2 \mathrm{var}\left[(z_i-\mu)^3\right]$$

$$+4g(\mu)g'(\mu)\left([g'(\mu)]^2 + g(\mu)g''(\mu)\right)E\left[(z_i-\mu)^3\right]$$

$$+g(\mu)g'(\mu)[g''(\mu)]^2 E\left[(z_i-\mu)^5\right] + 4g(\mu)[g'(\mu)]^2 g''(\mu)E\left[(z_i-\mu)^4\right]$$

$$+\frac{1}{2}[g''(\mu)]^2\left([g'(\mu)]^2 + g(\mu)g''(\mu)\right)\left(E\left[(z_i-\mu)^6\right] - \sigma^2 E\left[(z_i-\mu)^4\right]\right)$$

$$+2g'(\mu)g''(\mu)\left([g'(\mu)]^2 + g(\mu)g''(\mu)\right)\left(E\left[(z_i-\mu)^5\right] - \sigma^2 E\left[(z_i-\mu)^3\right]\right)$$

$$+\frac{1}{2}g'(\mu)[g''(\mu)]^3\left(E\left[(z_i-\mu)^7\right] - E\left[(z_i-\mu)^4\right]E\left[(z_i-\mu)^3\right]\right). \tag{6.4}$$

The calculation of the expectations for (a), (b) and (c) are provided below:

$$E[(a)] = E\left[(z_i-\mu)^6 - \sigma^2(z_i-\mu)^4 - (z_i-\mu)^2 E[(z_i-\mu)^4] + \sigma^2 E[(z_i-\mu)^4]\right]$$

$$= E[(z_i-\mu)^6] - \sigma^2 E[(z_i-\mu)^4] - \underbrace{E[(z_i-\mu)^2]}_{=\sigma^2}E[(z_i-\mu)^4] + \sigma^2 E[(z_i-\mu)^4]$$

$$= E[(z_i-\mu)^6] - \sigma^2 E[(z_i-\mu)^4]. \tag{6.5}$$

$$E[(b)] = E\left[(z_i - \mu)^5 - \sigma^2(z_i - \mu)^3 - (z_i - \mu)^2 E[(z_i - \mu)^3] + \sigma^2 E[(z_i - \mu)^3]\right]$$

$$= E[(z_i - \mu)^5] - \sigma^2 E[(z_i - \mu)^3] - \underbrace{E[(z_i - \mu)^2]}_{=\sigma^2} E[(z_i - \mu)^3] + \sigma^2 E[(z_i - \mu)^3]$$

$$= E[(z_i - \mu)^5] - \sigma^2 E[(z_i - \mu)^3]. \tag{6.6}$$

$$E[(c)] = E\left[(z_i - \mu)^7 - (z_i - \mu)^4 E[(z_i - \mu)^3]\right.$$
$$\left. -(z_i - \mu)^3 E[(z_i - \mu)^4] + E[(z_i - \mu)^3] E[(z_i - \mu)^4]\right]$$

$$= E[(z_i - \mu)^7] - E[(z_i - \mu)^4] E[(z_i - \mu)^3]$$
$$- E[(z_i - \mu)^3] E[(z_i - \mu)^4] + E[(z_i - \mu)^3] E[(z_i - \mu)^4]$$

$$= E[(z_i - \mu)^7] - E[(z_i - \mu)^4] E[(z_i - \mu)^3]. \tag{6.7}$$

Next, we do second-order Taylor Expansion of $H(z_i)$ at $z_i = \mu$, $i = 1, 2, ..., n$.

$$H(z_i) \approx H(\mu) + H'(\mu)(z_i - \mu) + \frac{1}{2}H''(\mu)(z_i - \mu)^2. \tag{6.8}$$

Take expectation on both sides,

$$E[H(z_i)] \approx H(\mu) + H'(\mu)\underbrace{E[z_i - \mu]}_{=0} + \frac{1}{2}H''(\mu)\underbrace{E\left[(z_i - \mu)^2\right]}_{=\sigma^2} = H(\mu) + \frac{1}{2}\sigma^2 H''(\mu). \tag{6.9}$$

$$\text{var}[H(z_i)] = E\left[\left(H(z_i) - E[H(z_i)]\right)^2\right]$$

$$\approx E\left\{\left(H'(\mu)(z_i - \mu) + \frac{1}{2}H''(\mu)(z_i - \mu)^2 - \frac{1}{2}\sigma^2 H''(\mu)\right)^2\right\}$$

$$= E\left\{\left(H'(\mu)(z_i - \mu) + \frac{1}{2}H''(\mu)\left[(z_i - \mu)^2 - \sigma^2\right]\right)^2\right\}$$

$$= [H'(\mu)]^2 \underbrace{E\left[(z_i - \mu)^2\right]}_{\sigma^2} + \frac{1}{4}[H''(\mu)]^2 \underbrace{E\left[\left((z_i - \mu)^2 - \sigma^2\right)^2\right]}_{(*)}$$

$$+ H'(\mu)H''(\mu)\left(E\left[(z_i - \mu)^3\right] - \sigma^2 \underbrace{E(z_i - \mu)}_{=0}\right)$$

$$=\left[H'(\mu)\right]^2 \sigma^2 + \frac{1}{4}\left[H''(\mu)\right]^2 \left(E\left[(z_i-\mu)^4\right]-\sigma^4\right) + H'(\mu)H''(\mu)E\left\{(z_i-\mu)^3\right\}. \quad (6.10)$$

For i.n.i.d. samples, let $E(z_i)=\mu_i$. The calculation procedure is pretty similar to that for the i.i.n. case. The only two modifications are that $E(z_i)=\mu_i$ instead of an identical mean value and individual $\text{var}\left[g_i^2(z_i)\right]$ and $\text{var}\left[H_i(z_i)\right]$ are summed to get $\sum_{i=1}^{n}\text{var}\left[g_i^2(z_i)\right]$ and $\sum_{i=1}^{n}\text{var}\left[H_i(z_i)\right]$.